\documentclass[onecolumn,showpacs,superscriptaddress,amsmath,amssymb,preprint,aps]{revtex4}

\usepackage{amsmath}
\usepackage{latexsym,amssymb}
\usepackage[mathscr]{eucal}
\usepackage{amsthm,amsxtra,amscd,upref}
\usepackage{layout,bm,dcolumn}
\usepackage{graphicx,color}
\usepackage{calc}

\newcommand{\cref}[1]{Chap.~\ref{#1}}

\begin{document}

\title{Diffuse scattering from the lead-based relaxor ferroelectric PbMg$_{1/3}$Ta$_{2/3}$O$_{3}$}
\author{A.~Cervellino}
\affiliation{Swiss Light Source, Paul Scherrer Institute, CH-5232 Villigen, Switzerland}
\affiliation{Laboratory for Neutron Scattering, ETHZ \& PSI, CH-5232 Villigen, Switzerland}
\author{S.N.~Gvasaliya}
\affiliation{Laboratory for Neutron Scattering, ETHZ \& PSI, CH-5232 Villigen, Switzerland}
\author{O.~Zaharko}
\affiliation{Laboratory for Neutron Scattering, ETHZ \& PSI, CH-5232 Villigen, Switzerland}
\author{B.~Roessli}
\affiliation{Laboratory for Neutron Scattering, ETHZ \& PSI, CH-5232 Villigen, Switzerland}
\author{G.M.~Rotaru}
\affiliation{Laboratory for Neutron Scattering, ETHZ \& PSI, CH-5232 Villigen, Switzerland}
\author{R.A.~Cowley}
\affiliation{Clarendon Laboratory, Department of Physics, Oxford University 
Parks Road, Oxford, OX1 3PU, UK}
\author{S.G.~Lushnikov}
\affiliation{Ioffe Physical Technical Institute, 194021 St. Petersburg, Russia} 
\author{T.A.~Shaplygina}
\affiliation{Ioffe Physical Technical Institute, 194021 St. Petersburg, Russia}
\author{M.-T.~Fernandez-Diaz}
\affiliation{Institut Laue-Langevin, 156X, 38042 Grenoble C\'{e}dex, France}
\date{\today}

\begin{abstract}
The relaxor ferroelectric PbMg$_{1/3}$Ta$_{2/3}$O$_{3}$ was studied by single-crystal neutron and 
synchrotron x-ray diffraction and its detailed atomic structure has been modeled in terms of 
static Pb-displacements that lead to the formation of polar nanoregions. Similar to the 
other members of the Pb-based relaxor family like PbMg$_{1/3}$Nb$_{2/3}$O$_{3}$ or 
PbZn$_{1/3}$Nb$_{2/3}$O$_{3}$ the diffuse scattering in the [H,0,0]/[0,K,0] scattering plane has 
a butterfly-shape around the (h,0,0) Bragg reflections and is transverse to the scattering vector 
for (h,h,0) peaks. In the [H,H,0]/[0,0,L] plane the diffuse scattering is elongated along 
the $<1,1,2>$ directions and is transverse to the scattering vector for (h,h,h) reflections. 
We find that a model consisting of correlated Pb-displacements 
along the $<1,1,1>$-directions reproduces the main features of the diffuse scattering in 
PbMg$_{1/3}$Ta$_{2/3}$O$_{3}$ adequately when the correlation lengths between the Pb-ion displacement vectors 
are longest along the $<1,1,1>$ and $<1,-1,0>$  and shortest along $<1,1,-2>$ directions. 
\end{abstract}
\maketitle
\section{Introduction}
Cubic perovskites $\rm AB'_xB''_{1-x}O_3$ with random occupation of the $\rm B$ site 
form a special class of ferroelectrics. In these materials the dielectric permittivity has a 
maximum around $\rm T_{max}$ that is not associated with any structural phase transition. Because the 
maximum of the dielectric constant is broad in temperature and also depends on the 
frequency, these materials are called "relaxor ferroelectrics". 
Pb$\mathrm{B}'_{1/3}\mathrm{B}''_{2/3}$O$_3$ ($\mathrm{B}'$=Zn,Mg; $\mathrm{B}'$=Nb,Ta) are 
typical relaxor ferroelectrics. 
To explain the broad maximum in the dielectric permittivity it was proposed by Smolenskii~\cite{smolenskii59} 
that fluctuations in $\rm B':B''$ ions lead to a coexistence of polar and non-polar regions in these 
materials as the temperature is reduced. Later Burns~\cite{burns} used the same approach 
to explain the temperature dependence of the refraction index that polar regions of nanometer size 
(PNR) form below a characteristic temperature, which for PbMg$_{1/3}$Nb$_{2/3}$O$_3$ (PMN) 
is $T_d\sim 620$~K. 

Extensive studies of the phonons in relaxor ferroelectrics by inelastic neutron scattering were unable to 
identify the soft mode that together with the Lyddane-Sachs-Teller relationship would explain the 
temperature dependence of the dielectric permittivity. Recently, however, it was shown 
that a quasi-elastic mode appears below the Burns temperature in PMN whose intensity increases with 
decreasing temperature and reaches a maximum at $T\sim$~380~K~\cite{gvasaliya_pmn_prb, hiraka}. 
The slow fluctuations associated with this scattering correspond to atomic motions correlated over a 
few unit cells only. In addition there is  strictly elastic diffuse intensity in PMN that appears 
at $T\sim$~420~K and indicates that static domains develop below that temperature~\cite{gvasaliya_pmn_jpco}.  
The intensity of the diffuse scattering increases with decreasing temperature and saturates below 
$T\sim$~100~K. These findings are in accord with a model of slowly fluctuating polar nano-regions 
in relaxor ferroelectrics that freeze with decreasing temperatures but do not undergo a phase 
transition to a long-range ferroelectric state due to the random-fields~\cite{cowley}.

Because the size of the polar nano-regions does not exceed a few hundreds \AA ngstr\"oms, these give rise to the 
presence of diffuse scattering in both x-ray and neutron spectra. The distribution of the 
diffuse scattering in reciprocal space was investigated by many groups. 
Vakhrushev {\it et al.}~\cite{vakhrushev95} found that the diffuse scattering in PMN is predominantly 
transverse to the scattering vector $\bm Q$ and that the intensity is weak near Bragg peaks with even 
parity ($2h, 2k, 2l$). Xu {\it et al.}~\cite{xu2004} investigated the distribution of diffuse scattering 
from PbZn$_{1/3}$Nb$_{2/3}$O$_{3}$-PbTiO$_3$ using x-ray techniques and found that in the 
(h,k,0) plane the diffuse intensity consists 
either of elongated streaks perpendicular to the scattering vector $\bm Q$ (e.g. around (1,1,0)) or 
has the shape of a "butterfly" (e.g. around (1,0,0)). Although there is still no complete understanding 
of the shape and size of the PNR in relaxor ferroelectrics, it has been recognized that it is mainly 
correlated Pb-ions displacements that produce the diffuse scattering and are at the origin of the polar 
nano-regions~\cite{welberry2006}. While the direction of the Pb-displacements is difficult to 
obtain unambiguously from powder diffraction measurements, the average amplitude of the atomic 
displacements generally agrees with the results of modeling the diffuse-scattering. For example early 
powder neutron diffraction measurements from PMN by Bonneau {\it et al.}~\cite{bonneau} showed that Pb 
ions are displaced from the (0,0,0) position below $T\sim$800~K and that the magnitude of these 
displacements increases with decreasing temperature. At T=5~K the Pb ions are statistically displaced 
by $\sim 0.36$~\AA { } along the $<1 1 1>$ directions. The probability density function of Pb in PMN 
was determined~\cite{zhukov,vakhrushev2002} and it is found that while at room temperature Pb is 
isotropically displaced, the Pb potential has a double-well shape at low temperature.

PbMg$_{1/3}$Ta$_{2/3}$O$_{3}$ (PMT) is a typical relaxor ferroelectric. As for PMN the 
average crystal structure is  cubic at all temperatures. The real part of the dielectric 
permittivity has a maximum at $T\sim$170~K and at a frequency of $\nu$=10~KHz. 
Powder neutron diffraction~\cite{gvasaliya2004} has shown that the amplitude of the Pb displacements 
grows rapidly below the Burns temperature that is $\sim570$~K for PMT~\cite{markovin1992}. The 
Pb-displacements give rise to neutron diffuse scattering 
whose temperature dependence correlates with the temperature dependence of the 
atomic displacements well. Because of the similarities existing between the physical properties of PMT 
with those of PMN and PbZn$_{1/3}$Nb$_{2/3}$O$_{3}$ (PZN), we decided to investigate the detailed 
atomic structure by neutron and synchrotron radiation. 
The diffuse scattering of PMT in the [H,0,0]/[0,K,0] plane is transverse to 
the scattering vector for the (h,h,0) Bragg reflections and has a butterfly-shape near 
the (h,0,0) peaks. In the [H,H,0]/[0,0,L] plane the diffuse scattering of PbMg$_{1/3}$Ta$_{2/3}$O$_{3}$ 
is elongated along $<1,1,2>$ directions which indicates that the motion of the Pb ions 
have a short correlation length along that direction. 

In the following we will denote $\bm{Q}$ the general momentum transfer vector 
($Q=|\bm{Q}|=4\pi\sin(\theta)/\lambda$, with $2\theta$ the scattering angle and $\lambda$ 
the incident wavelength). The reciprocal lattice vectors will be denoted by $\bm{\tau}$, and 
the relative momentum transfer vector by $\bm{q}=\bm{Q}-\bm{\tau}$. 
\section{Experimental Results}
\subsection{Single crystal neutron diffraction}

We performed a high-quality neutron single crystal experiment from a PMT sample on D9 at ILL. 
The diffractometer is located on the hot source and allows access
to Bragg reflections at large $Q$ values. The short neutron wavelength of 0.513~\AA, combined with the small crystal size 
($1.6\times 1.1\times 0.8$~mm) helped also to minimize extinction problems. 
Bragg intensities were collected at $T=20$~K and $T=300$~K,
respectively. At room temperature 632 reflections were measured up to 
$Q=4\pi\sin(\theta)/\lambda=15.82$~\AA $^{-1}$
and 612 ($Q\leqslant 19.45$~\AA $^{-1}$) at $T$=20 K, with maximum Miller index 11. A symmetry test showed
that the crystal symmetry is cubic at both temperatures, R$_{eq}$=3.35\% for the 300~K data set 
and R$_{eq}$=5.92\% at 20~K. 

\begin{figure}[ht]
  \includegraphics[width=0.99\textwidth]{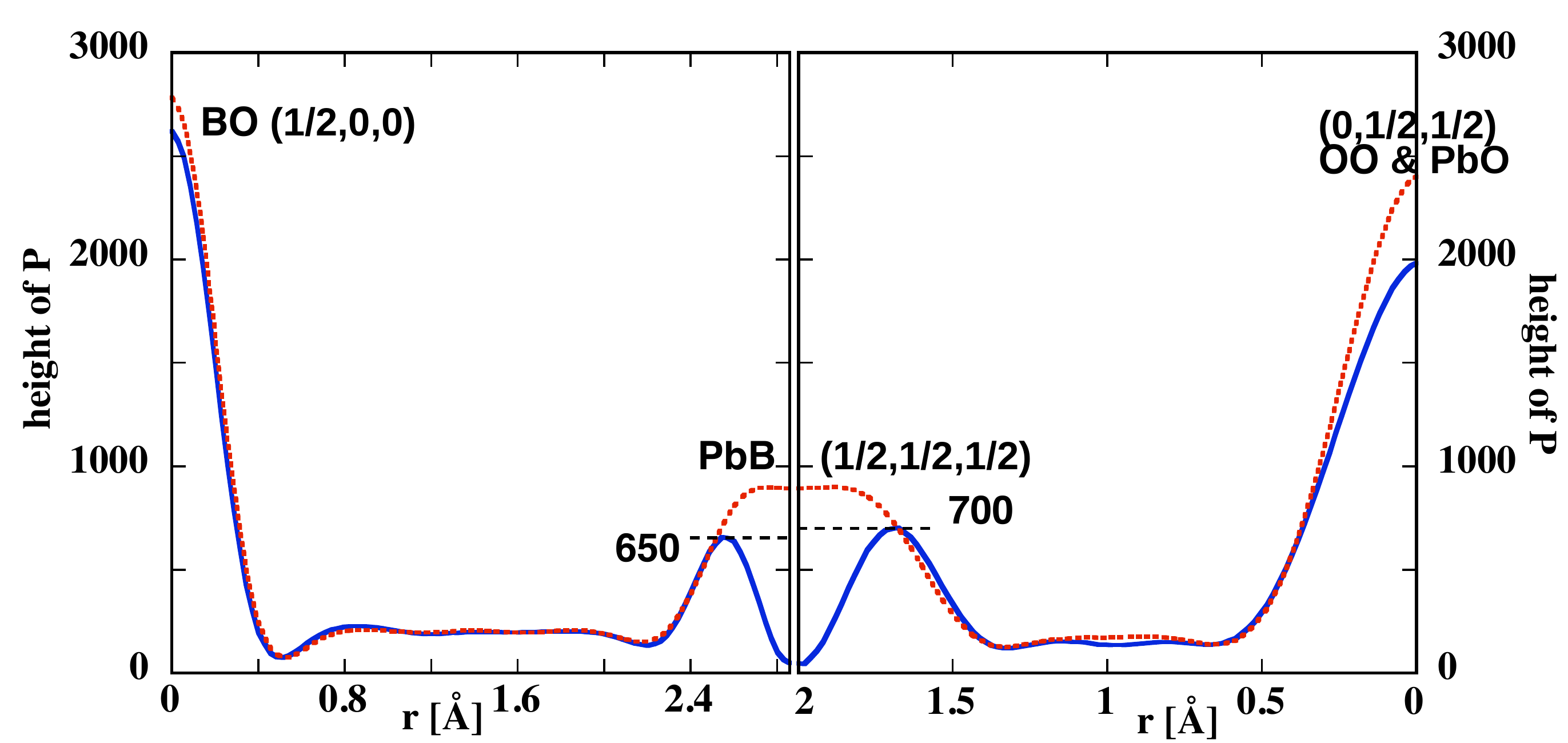}
  \caption{Linear cuts through the Patterson maps at 20 K (blue) and 300 K (red) along the $<0,1,1>$ ({\it left}) 
and $<1,0,0>$ ({\it right}) directions. Endpoints in crystal coordinates are shown for clarity, 
together with the atom pairs corresponding to the relevant Patterson map peaks.
           }
\label{fig1}
\end{figure}

The cubic symmetry describes the average periodic structure only. 
Atomic displacements break this symmetry locally, however the global symmetry is retained through 
the formation of domains. So, PMT must be described as a perturbed periodic 
structure. Whereas the aim of this paper is to clarify the nature of the perturbation, the 
diffraction data contain information about the size of the displacements that will be used 
as a constraint for the analysis of the diffuse scattering patterns. To analyze the atomic displacements we used 
the Patterson function $P(\bm{u})$ that is calculated through the Fourier transform of the observed Bragg
intensities:
\begin{equation}
\label{patterson}
P(\bm{u}) =
\mathop{\sum}_{\bm{\tau}}I_{\bm{\tau}}\cos({\bm{\tau}}\cdot\bm{u})=
\frac{1}{v}\int {\mathrm{d}}^3\bm{r}\rho(\bm{r})\rho(\bm{u}+\bm{r})
\end{equation}
where $\bm{\tau}$ are the Bragg reflections, $I_{\bm{\tau}}$ the relevant observed intensities, 
$\rho(\bm{r})$ the scattering density and $v$ is the unit cell volume. 
In order to properly set the zero of $P(\bm{u})$ the value of $I_{000}$ - calculated on the same scale after the structure refinement below described -  
has been added to the list of the observed intensities. 
Eq.~\ref{patterson} shows that the Patterson function is the self-convolution of the scattering 
density, so the position of the peaks in $P({\bm{u}})$ gives direct access to the interatomic distances
${\bm{u}}$ \cite{giacovazzo}. Cuts through the Patterson maps across the main interatomic peaks are shown in Fig.~\ref{fig1}.  
The most pronounced characteristic in the cut through the Patterson map shown in Fig.~\ref{fig1} is visible 
in the Pb-B interatomic peaks. At room temperature the Pb-B average length is $\sim$2.5~\AA { } with a distribution 
of $\sim$0.3~\AA. At low temperature the Pb-B length shows a maximum away from the expected distance in the 
perovskite structure. 
As B-O peaks do not show anything similar, we must attribute this feature to displacements of Pb atoms from 
their average position. At room temperature the Pb atoms lie in a spherical region about their nominal positions. At 
$T=20$~K, the Pb displacements are more defined and Pb atoms lie on an isotropic spherical shell. 
This can be described by a 3-D double-well-shaped potential defining the 
Pb displacements. 

Moreover, a least-square fit to the diffraction data was done using the programs {\texttt{jana2000}}~\cite{jana}. 
The refined atomic parameters are summarized in Table I.
In agreement with the analysis of Vakhrushev {\it et al.}~\cite{vakhrushev2002} for PMN, 
the structural refinement at $T=20$~K 
indicates simultaneous displacements of the Pb ions along all directions by  0.34(2)\AA. 
The displacements of the O ions are modeled by anisotropic temperature factors. The result of the 
refinement gives that the 
O thermal ellipsoids have a larger component in the $<110>$-plane. 
We also observe that the oxygen-related Patterson peaks 
increase and get sharper with increasing temperature, which is direct evidence for the anomalous decrease
of the thermal displacement parameters of the oxygen atoms observed in a previous powder
diffraction experiment~\cite{gvasaliya2004}. So, while we cannot disregard 
the possibility of O displacements correlated with the Pb displacements, 
this is not a dominant feature. 
We could not obtain from the diffraction evidence about the anisotropy of the displacements, 
although the spatial resolution would be in principle sufficient for this purpose. 
We attribute this limitation to either the averaging over many symmetry-equivalent PNR domains
 or to a large fraction of thick domain walls (or non-polar regions~\cite{smolenskii59}) between PNR, that are not ordered.
Furthermore even in ordered PNR the displacements are only broadly collinear. However, 
the results obtained from the diffraction data allow us to restrict the analysis of the diffuse scattering 
to Pb displacements only. 
\subsection{X-ray diffuse scattering}
\label{SSX}

The diffuse scattering was measured in a single crystal of PMT at the Swiss-Norwegian Beam 
Line (SNBL) at the ESRF. The measurements were performed in transmission geometry with the wavelength $\lambda=0.71076$~\AA{} 
in a broad temperature range between $T=100$~K and 450~K. Systematic maps of 
scattered intensity were collected by a MAR-345 imaging plate detector with the single crystal being 
rotated in steps of 0.5$^0$ around the vertical axis. From this data the distribution of the 
diffuse intensity in the [H,0,0]/[0,K,0] and [H,H,0]/[0,0,L] scattering planes could be reconstructed, using the program 
{\tt{CrysAlis}} by Oxford Diffraction to merge the measured maps. 

\begin{figure}[ht]
  \includegraphics[width=0.49\textwidth]{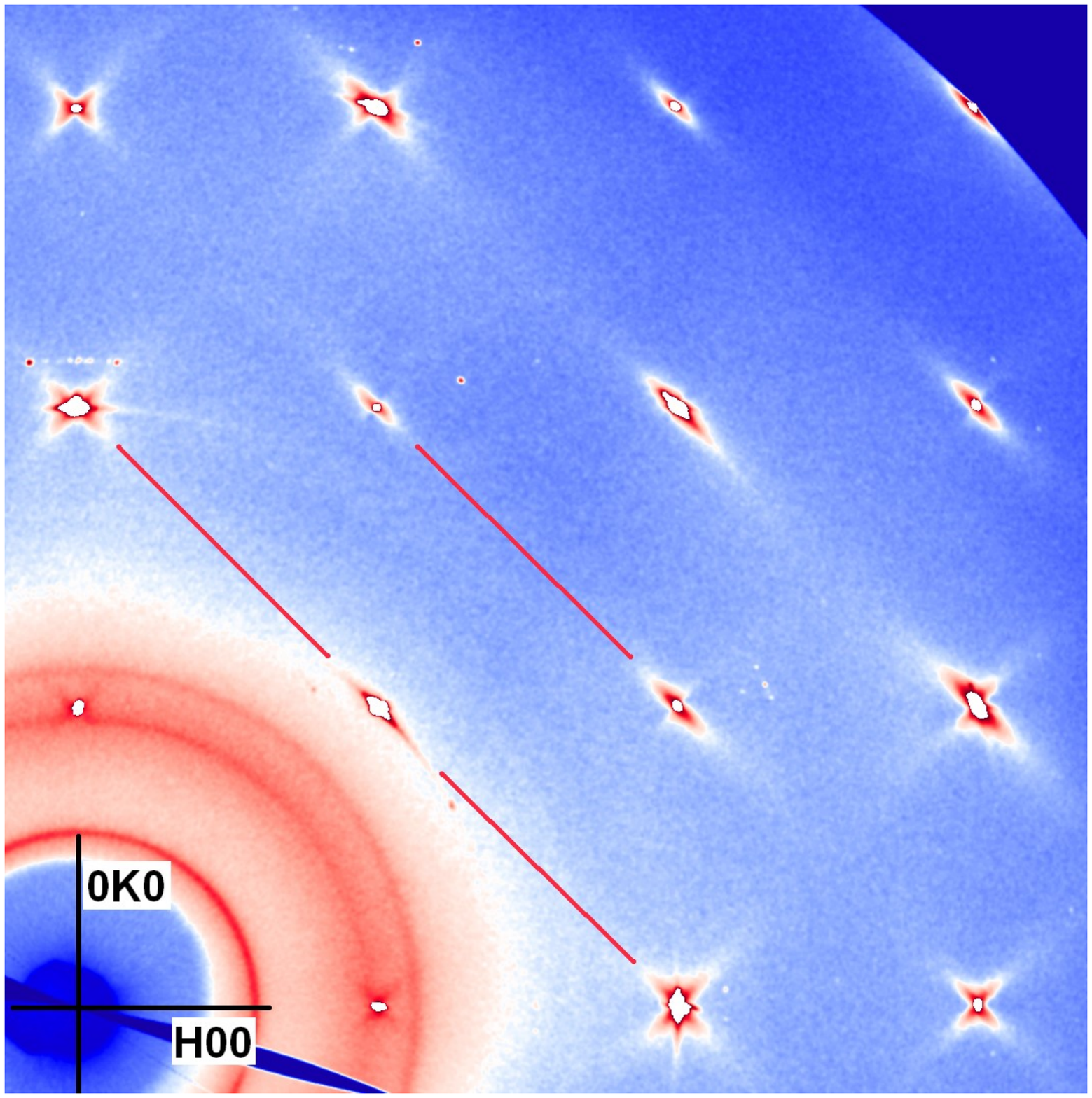}
  \includegraphics[width=0.49\textwidth]{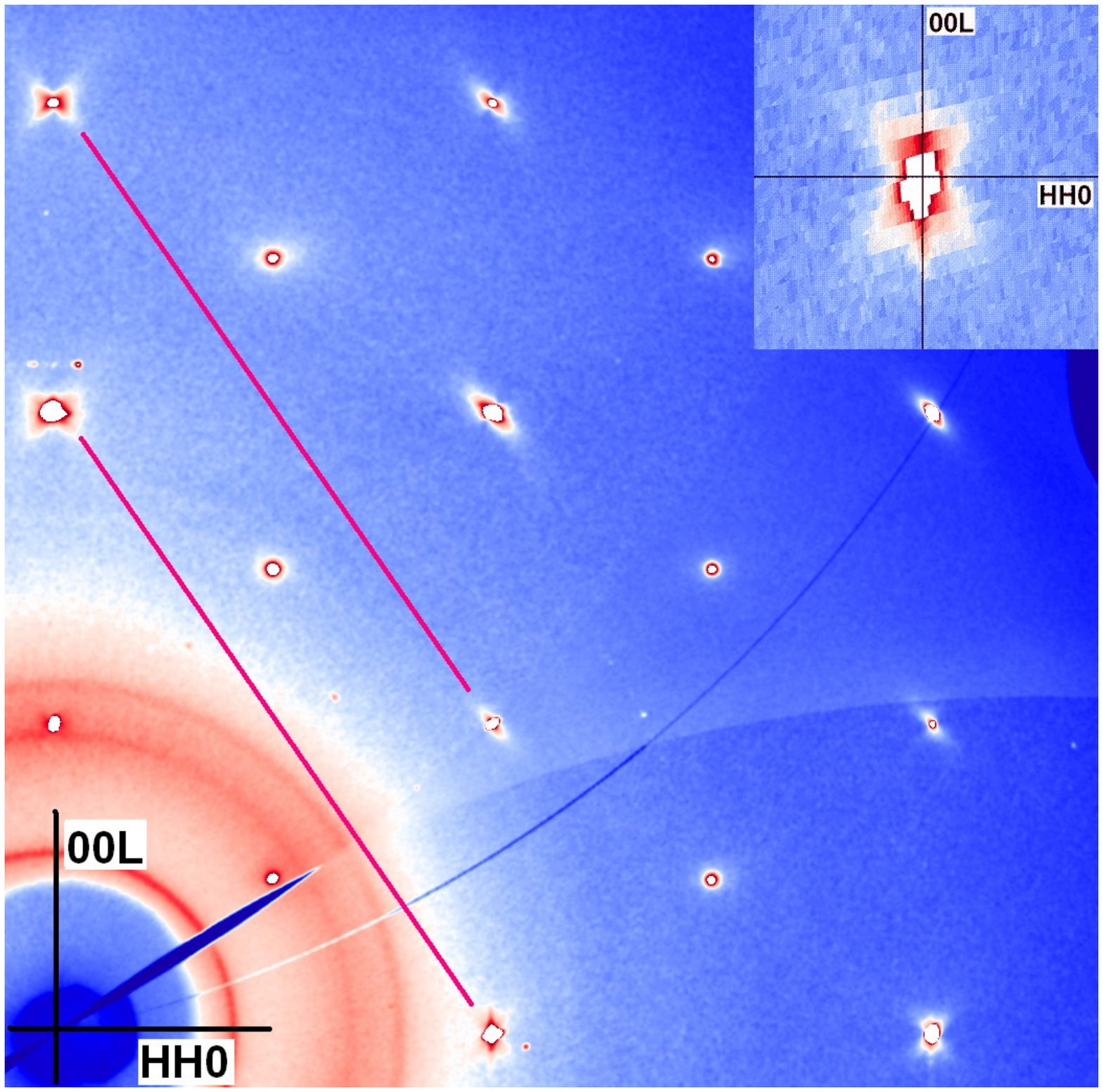}
  \caption{a) False-color reconstructed image of experimental diffuse scattering intensities from PMT 
in the [H,0,0]/[0,K,0] scattering plane. 
The direction of streaks of diffuse intensity are shown by a solid 
line and extend along the $<1,1,0>$-type directions.  
b) False-color image of experimental intensities from PMT in the 
[H,H,0]/[0,0,L] scattering plane. 
The direction of streaks of diffuse intensity are shown by a solid 
line and extend along the $<1,1,2>$-type directions.
The intensity spots visible at the 
$\frac{1}{2}(h,h,l)$ positions are due to partial order in the B-sublattice.
The data was taken at T=175~K. The inset to Fig.~\ref{fig2}b shows the 
distribution of the diffuse scattering around the (2,2,0) Bragg peak. 
The distance from the center of the inset to the boundaries along the 
$<H,H,0>$ and $<0,0,L>$ directions is 0.78{\AA $^{-1}$}. Note that the 
diffuse scattering is elongated along the $<1,1,2>$ direction.
}     
\label{fig2}
\end{figure}

Figure~\ref{fig2} shows the reconstruction of the scattered intensity distribution in the [H,0,0]/[0,K,0] 
and [H,H,0]/[0,0,L] scattering planes at $T=175$~K.  
As known from previous neutron measurements, diffuse scattering is particularly strong at this temperature. 
The diffuse scattering in PMT has a similar "butterfly"-shaped distribution as found in PMN around the (2,0,0) Bragg reflection 
and within the [H,0,0]/[0,K,0] scattering plane as shown in Fig.~\ref{fig2}a. It can be observed that the 
diffuse scattering is elongated along the four equivalent $<1,1,0>$ directions around these zone centers, 
while the diffuse scattering is essentially transverse to the scattering vector $\bm Q$ around the (2,2,0) 
Bragg peak. Comparing these measurements with the diffuse scattering obtained when the single crystal is 
aligned in the [H,H,0]/[0,0,L] plane shows firstly that the diffuse scattering is extended along the 
$<1,1,2>$ directions around the (2,0,0) Bragg peak. Secondly, it can be observed in Fig.~\ref{fig2}b 
the diffuse scattering around the (2,2,0) Bragg peak has four wings of intensity that extend along the 
$<1,1,2>$ directions and no diffuse intensity is visible along the $<1,1,0>$ directions. This last observation 
shows that the intensity of the diffuse scattering depends on the direction of the atomic displacements 
relative to the orientation of the scattering vector $\bm Q$. The same effect was observed in 
PMN~\cite{xu2004,you}, where the diffuse scattering was modeled by introducing a polarization factor of 
the form $({\bm Q}\cdot {\bm u})^2$. Although this polarization factor has the same form as for phonon 
scattering, it is also present in the cross-section for diffuse x-ray and neutron scattering if atoms 
are displaced from the average position in the unit cell. In the latter case, the diffuse intensity varies 
like $({\bm Q}\cdot {\bm u})^2$ with $\bm u$ the direction of the atomic displacements~\cite{kirvoglaz}. 
It is this polarization factor that causes the absence of the diffuse intensity when the displacement 
disorder is perpendicular to the scattering vector $\bm Q$ as we observe for PMT in the vicinity of 
the (1,1,0) Bragg peak. As the diffuse scattering in PMT at low temperatures was found to be truly 
elastic from neutron diffraction measurements~\cite{gvasaliya2003}, it must originate from static 
random displacements, rather than from the condensation of a transverse optic phonon mode. 
\section{Model of the diffuse scattering in PMT}
To analyze the diffuse scattering in PMT we introduce a model that takes into account the shape 
of the polar nanoregions as well as the direction and amplitude of the displacements. We define therefore 
a shape function $S({\bm r})$ for a PNR. 
The autocorrelation function 
$G({\bm r})$ of $S({\bm r})$ is 
\begin{equation}
G({\bm r})=\int {\mathrm{d}}^3 {\bm r} S({\bm r}')S({\bm r}-{\bm r}'),
\end{equation} 
and its Fourier transform
\begin{equation}
{\tilde G}({\bm{Q}})=\int d^3 {\bm r}G({\bm r})\exp(- \mathrm{i} \bm{Q}\cdot{\bm r}).
\end{equation} 
For the case of a periodic crystal of limited dimensions described by $S({\bm r})$, 
the diffracted intensity is given by 
\begin{equation}
I({\bm Q})=|F({\bm Q})|^2\sum_{\bm \tau} {\tilde G}({\bm Q}-{\bm \tau}),
\end{equation}
where $F({\bm Q})=\sum_j \rho_j({\bm r})\exp(-{\mathrm{ i}} {\bm Q}\cdot{{\bm r}_j})$ 
is the structure amplitude of one unit cell and $\rho_j({\bm r})$ the scattering density of 
atom $j$; ${\bm \tau}$ is a reciprocal lattice vector.

For a disordered crystal the scattering density $\rho({\bm r})$~\footnote{We neglect here Debye-Waller 
and atomic form factors.} can be decomposed into an average atomic density 
$\rho_a({\bm r})\propto\delta({\bm r}-{\bm r}_j)$ and a random density 
$\rho_b({\bm r})\propto\delta({\bm r}-{\bm r}_j-{\bm u})-\delta({\bm r}-{\bm r}_j)$, so that 
the atomic distribution in the crystal is described by $\rho({\bm r})=\rho_a({\bm r})+\rho_b({\bm r})$. 
The average atomic structure contributes to the Bragg reflections and its scattered intensity is 
vanishingly small away from the reciprocal lattice vectors. In that case it is justified 
to neglect interference effects and to make the assumption that 
$|F({\bm Q})|^2\approx|F_a({\bm Q})|^2+|F_b({\bm Q})|^2$. In this approximation the total scattered 
intensity consists of the superposition of Bragg peaks with broad diffuse components centered 
around the lattice reciprocal vectors.
\subsection{The form factor}
The diffuse scattering measured in PMT was modeled by assuming that only the Pb ion is displaced 
from its average  position (${\bm r}_{\mathrm{Pb}}=(0,0,0)$ in the 
$Pm{\bar 3}m$ space group). If we suppose that the 
Pb atom is displaced  by a vector $\bm u$, we obtain for the random density 
\begin{equation}
F_b({\bm Q})=
b_{\mathrm{Pb}}[{\rm e}^{ -{\mathrm{ i}} {\bm Q}\cdot({\bm r}_{\mathrm{Pb}}+{\bm u})}-{\rm e}^{-{\mathrm{ i}}  
{\bm Q}\cdot{\bm r}_{\mathrm{Pb}}}],
\end{equation}  
where $b_{Pb}$ is the form factor (including Debye-Waller) of Pb ion 
in the unit cell. As in our simplified model the cubic unit cell of PMT contains only one 
ion (Pb) that is displaced  from its average site, 
the form factor of the diffuse intensity reduces to 
\begin{equation}
|F_b({\bm Q)}|^2=2b^2_{Pb}[1-\cos( {\bm Q}\cdot{\bm u})]=4b^2_{Pb}\sin^2( {\bm Q}\cdot{\bm u}/2).
\label{vec}
\end{equation}
For small values of ${\bm Q}\cdot{\bm u}$ eq.~\ref{vec} is $\propto({\bm Q}\cdot{\bm u})^2$ which is the 
form used in refs.~\onlinecite{xu2004,you}.

It is instructive to give here the $\bm Q$-dependence of the diffuse scattering when Pb ions in 
the {\it average} structure are located on a spherical shell of radius $u$, {\it i.e.}
\begin{equation}
\rho_a({\bm r})=b_{Pb}\frac{\delta(|{\bm r}-{\bm r}_{\mathrm{Pb}} |-u)}{4\pi u^2},
\end{equation} 
as suggested by neutron diffraction in both PMN refs~\onlinecite{bonneau,vakhrushev2002} and PMT (\emph{cf.} Sec.~\ref{SSX} and ref.~\onlinecite{gvasaliya2004}).
In that case, we obtain 
\begin{equation}
F_b({\bm Q})=b_{\mathrm{Pb}}{\rm e}^{-{\mathrm{ i}} {\bm Q}\cdot{\bm r}_{\mathrm{Pb}}}
\left[{\rm e}^{-{\mathrm{ i}}{\bm Q}\cdot{\bm u}}-\frac{\sin( Qu)}{ Qu}
\right],
\end{equation}
and the scattered diffuse intensity is modulated by a structure factor
\begin{equation}
|F_b({\bm Q})|^2=b^2_{\mathrm{Pb}}
\left[1+\left(\frac{\sin(Qu)}{ Qu}\right)^2-2\frac{\sin( Qu)}{ Qu}\cos({\bm Q}\cdot{\bm u})\right],
\label{shell}
\end{equation}
whose first nonzero term in the small- ${\bm Q}\cdot{\bm u}$ expansion is still $\propto({\bm Q}\cdot{\bm u})^2$.

Finally it should be pointed out that the dependence of the diffuse intensity in the case of 
atomic displacements differs from  Huang scattering that arises for small concentrations of defects. 
In this case, the displacements distribution has spherical symmetry around a defect and the scattering is proportional 
to $I_{Huang}\propto {(Q\cos(\phi))^2}/{q^2}$, where $\phi$ is the angle between $\bm Q$ and the 
wave-vector $\bm q$. Huang scattering is present around all Bragg reflections and its intensity is enhanced 
along the scattering vector ${\bm Q}$ while the diffuse intensity vanishes when ${\bm Q} \perp {\bm q}$. 
Inspection of the diffuse scattering measured in PMT at $T=175$~K shows that the planes of zero intensity 
are not always perpendicular to $\bm Q$. For example Fig.~\ref{fig2} shows that the diffuse scattering 
measured in PMT around the (1,1,0) Bragg position with the crystal being oriented with an $[0,0,1]$ 
axis vertical is 
elongated in the direction transverse to the scattering vector and is zero along $\bm Q$. Both 
results are contrary to what would be expected for Huang scattering. We note that it 
was found that weak diffuse scattering is still present in PMN above the Burns temperature, 
with, however, a different distribution of intensity in reciprocal space. At high temperature 
this residual diffuse scattering could mainly arise from Huang scattering and the chemical disorder
on the B-site. On the other hand correlated atomic displacements,  
which give rise to the strong diffuse scattering, disappear above the Burns temperature~\cite{gehring_2009}.
\subsection{The choice of the line-shape}
The line-shape of the diffuse scattering depends 
upon the shape function $S(\bm r)$ that describes the form and the size of the polar nanoregions. 
In the simplest assumption, if Pb displacements were to be perfectly aligned within any 
sharply defined geometric region, there would be characteristic intensity modulations, which have not  
been observed. Moreover, Monte-Carlo simulations of the diffuse scattering in PMN have 
shown that the domain wall structure of the polar nanoregions is complicated~\cite{pasciak2007} which 
makes it difficult to describe the shape-function with a simple analytical function. However, 
the autocorrelation of the shape $G({\bm r})$ can be described by an analytical function. 
In fact, shape in this case is defined in a stochastic way. 

Hereafter we choose to describe $G({\bm r})$ by decaying exponentials with an anisotropic decay 
length. Depending on the detailed symmetry of the decay length, this leads to power-law 
decay of the diffuse scattering. In fact, it has been demonstrated in PMN that the 
line-shape of the diffuse scattering generally follows a $q^{-\alpha}$-dependence, 
with $\alpha>2$ and weakly depending on $T$, becoming higher at room temperature~\cite{you2000}. 
Others (Chetverikov {\it et al.}~\cite{chetverikov} found that 
$\alpha$ varies between 1.5 and 2.6 as a function of the temperature. The 
"pancake" model of Xu {\it et al.}~\cite{xu2004} assumes that PNR have 
cylindrical symmetry, which yields an exponent $\alpha=2$ along the axis, $\alpha=3$ 
in the orthogonal plane, and higher along general directions~\cite{antonio}.  
In the following we will assume a more general (orthorhombic) anisotropy of the 
autocorrelation function, keeping its exponential character:
\begin{eqnarray}
G({\bm r})&=&{\mathrm{e}}^{-{\bm w}_1\cdot{\bm r}/L_1}{\mathrm{e}}^{-{\bm w}_2\cdot{\bm r}/L_2}{\mathrm{e}}^{-{\bm w}_3\cdot{\bm r}/L_3};
\nonumber\\
{\tilde G}({\bm q}) & = & C 
  \frac{1}{1+(L_1{\bm w}_1\cdot{\bm q})^2}\frac{1}{1+( L_2{\bm w}_2\cdot{\bm q})^2}\times\\
&&\times\frac{1}{1+( L_3{\bm w}_3\cdot{\bm q})^2}.  \nonumber
\end{eqnarray}
$L_1$, $L_2$, $L_3$ correspond then to the average dimensions of the polar nanoregions along the mutually 
orthogonal directions defined by the unit vectors ${\bm w}_i$ (i=1,2,3), $C$ is an inessential constant. 
This is consistent with power-law decay with $2<\alpha<4$ along special directions ($\alpha\leqslant 6$ in general). 
\section{Interpretation of the diffuse scattering in PMT}
In order to model the diffuse scattering in PMT, the direction of the 
static displacements has to be known. There is however no consensus about the direction 
of the Pb displacement in the literature. In Xu's analysis~\cite{xu2004} of the diffuse 
scattering in PMN, the polarization (displacement) direction is parallel to $<1,-1,0>$. 
In contrast using neutron scattering and a pair distribution function analysis, 
Jeong {\it et al.}~\cite{jeong2005} found direct evidence that displacements in PMN have 
rhombohedral symmetry, which suggests that the Pb atoms are displaced along the $<1,1,1>$ 
direction.
Initially we assume that ionic displacements $\bm u$ are along the $<1,-1,0>$ equivalent directions, and choose 
the unit vectors that define the dimensions of the PNR 
as ${\bm w}_1$=[1,1,0]/$\sqrt{2}$, ${\bm w}_2$=[1,-1,0]/$\sqrt{2}$ and ${\bm w}_3$=[0,0,1], 
respectively. In this model, there are six domains which correspond to the symmetrically equivalent 
$<1,1,0>$ directions in the cubic structure. When $L_1\ll L_2=L_3$, the model is 
equivalent to the "pancake" model of the PNR of ref.~\onlinecite{xu2004}. 
This model accounts properly for the distribution of the diffuse scattering in 
PMT in the (1,0,0) Brillouin zones with $<0,0,1>$ perpendicular to the scattering plane when 
$L_1\sim$20~\AA {} and $L_2=L_3\sim$~100~\AA. In the [H,H,0]/[0,0,L] scattering plane, the diffuse 
scattering has the shape of a "butterfly" with wings along the $<1,1,1>$ directions that is reproduced 
by the calculations. We find, however, that this model does not reproduce the distribution of 
diffuse scattering  intensities around the (2,2,2) Bragg peak properly. In fact, the measurements clearly show that diffuse scattering 
is mainly transverse to the scattering vector $\bf Q$, as shown in Fig.~\ref{fig2}b, while this model 
yields additional streaks along $\bf Q$ that cannot be eliminated alone by adjusting the correlation lengths. 
We then concluded that the transverse streak is cancelled because of the  form factor $\propto(\bm{Q}\cdot\bm{u})^2$. 
In addition, inspection of the diffuse intensity map in PMT reveals that in the 
[H,H,0]/[0,0,L] scattering plane the streaks of diffuse scattering around the (0,0,l) 
Bragg peaks are extended along the $\pm <1,1,\mp~ 2>$ directions which 
indicates that the length of the polar nanoregions must be short along these directions. 
The simplest way of building PNR with e.g. a short $<1,1,-2>$ extension is to choose 
${\bm w}_1=1/\sqrt{2} [1,-1,0]$, ${\bm w}_2$=$1/\sqrt{3}[1,1,1]$ and ${\bm w}_3$=$1/\sqrt{6}$[1,1,-2]. 
In this case there are 12 possible PNR domains that will contribute to the patterns 
of diffuse intensity. The diffuse scattering distribution was calculated for displacements $\bm u$ along these directions and the best agreement between observed and calculated diffuse intensity patterns 
could be obtained by choosing ${\bm u}$ along $<1,1,1>$. For the other 
directions of atomic displacements  the distribution of intensity wings around 
some of the Bragg peaks were incorrect.

\begin{figure}[ht]
  \includegraphics[width=0.49\textwidth]{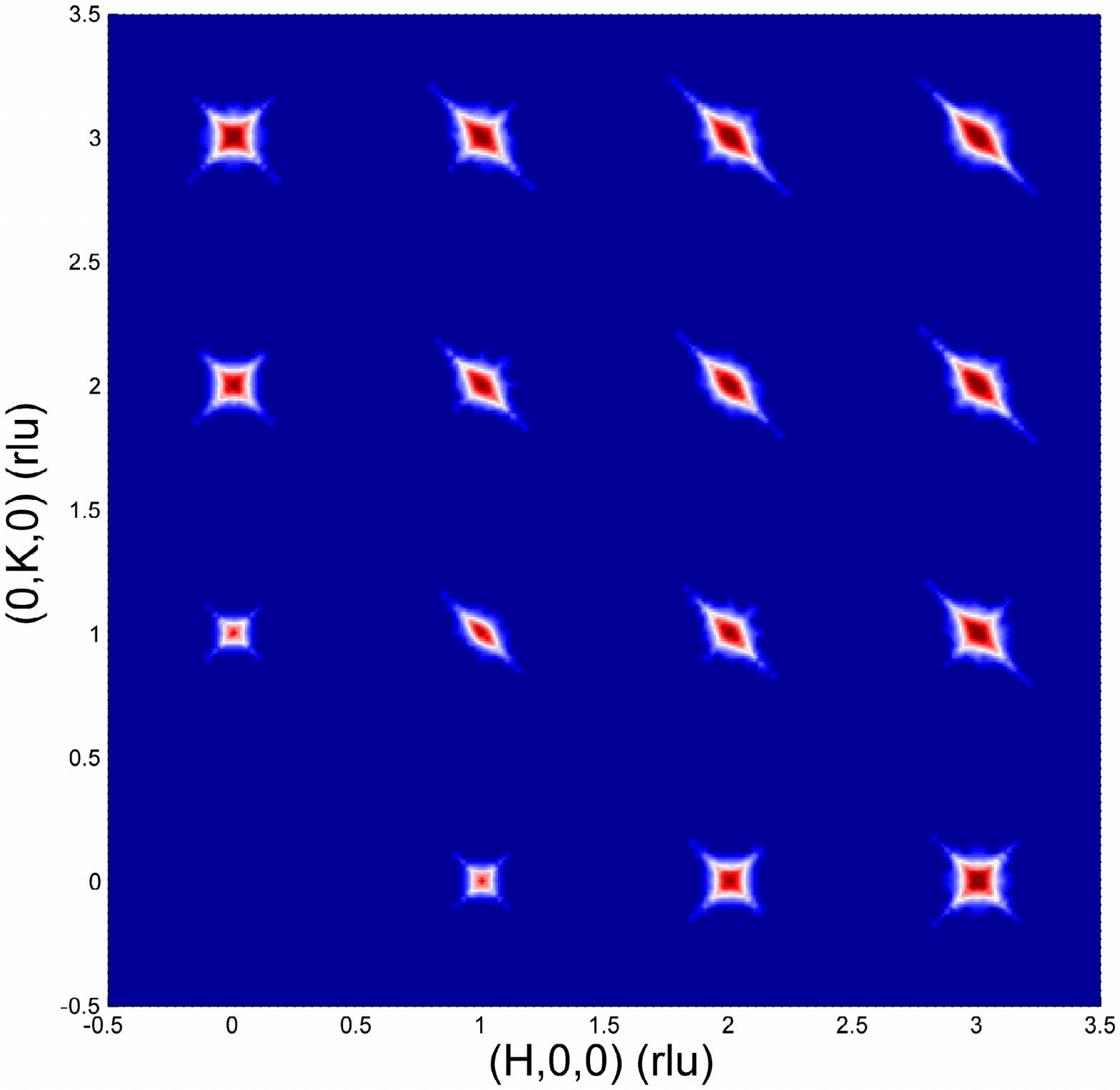}
  \includegraphics[width=0.49\textwidth]{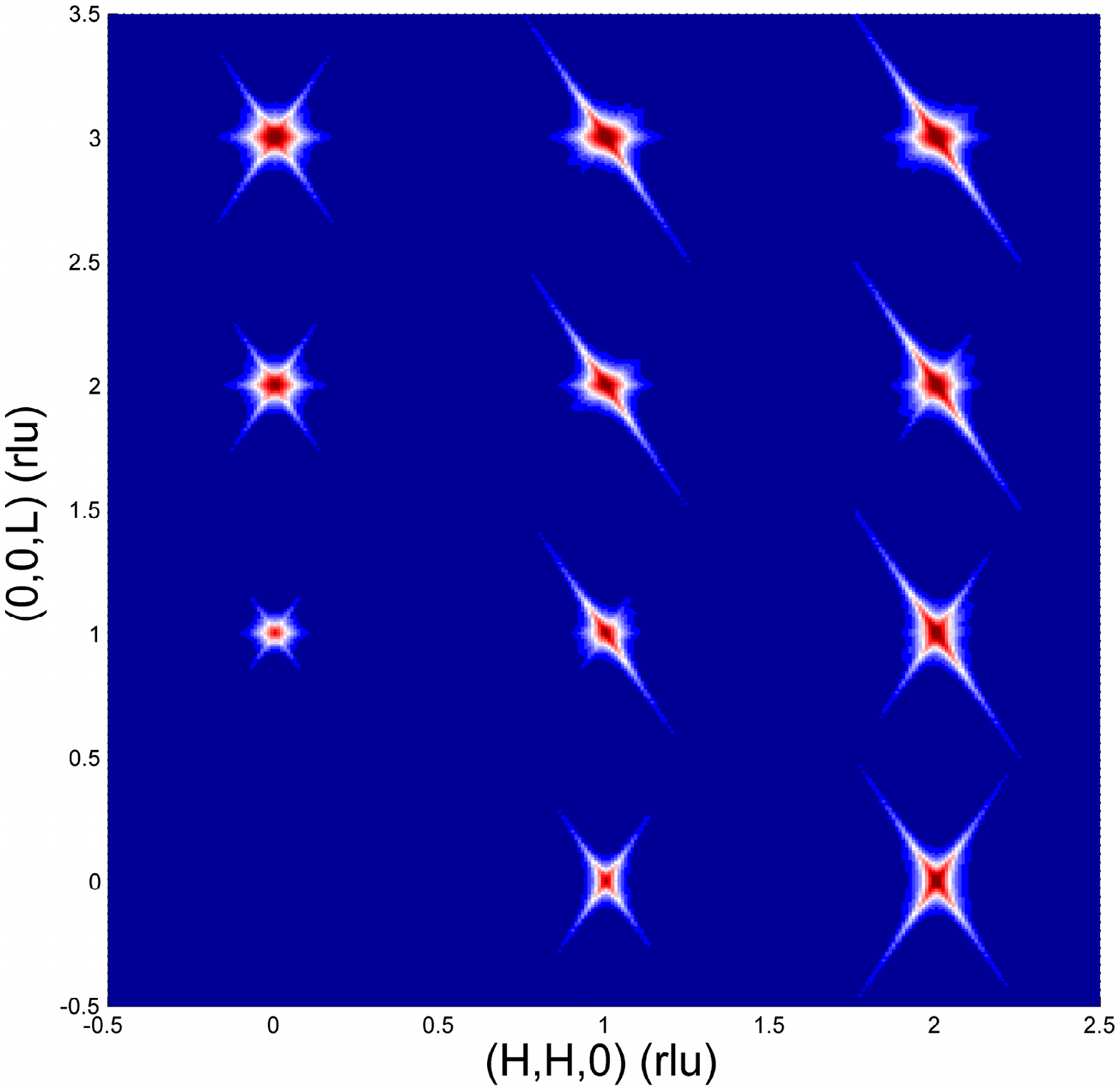}
  \caption{False-color images of calculated diffuse scattering intensities from PMT 
           a) in the [H,0,0]/[0,K,0]- and b) in the [H,H,0]/[0,0,L]-scattering planes. 
           The intensities are given in a logarithmic scale. 
           }
\label{fig3}
\end{figure}

In the previous section we showed that the model with the Pb determined the direction of the 
most probable Pb-displacements in PMT as $<1,1,1>$, the dimensions of the parallelepiped could be 
estimated from the width of the diffuse intensity away from the Bragg peaks. At $T=175$~K, 
a reasonable estimation yields $L_1\sim$50~\AA; $L_2\sim$30~\AA {} and 
$L_3\sim$4~\AA. Calculated diffuse scattering maps are shown in Figs.~\ref{fig3}a and \ref{fig3}b 
and can be compared with the experimental data presented in Fig.~\ref{fig2}.
\section{Discussion and conclusions}
Our main results can be summarized as follows:
\begin{itemize}
\item[i)\quad] Patterson analysis provides direct evidence that the potential at Pb sites 
develops a double-well structure at low temperature causing Pb ions to be 
distributed on a shell with radius $\approx 0.3$~\AA. 
The observed extinction of some diffuse streaks is in agreement with the form of 
the structure factor given by eqs~\ref{vec},\ref{shell}. In addition it should
be possible to measure the change in the Pb potential with temperature by computing the
diffuse scattering intensities as a function of scattering vector $\bf Q$. As the displacements are
important ($u \approx 0.2 ... 0.3$\AA), this would require recording the 
distribution of the diffuse scattering intensity at large scattering vectors.
Having Pb ions displaced along the $<1,1,1>$ direction is consistent with the analysis
of the local structure of relaxor ferroelectrics by PDF \cite{jeong2005} and with the observations of a
rhombohedral phase transition that can be induced by application of an electric field \cite{calvarin}.
Also recent dielectric measurements in PMN doped with PbTiO$_3$ in applied electric field have shown that
the spontaneous polarization in the polar nanoregions is along $<1,1,1>$ \cite{li2008}. 
\item[ii)\quad ] Describing the structure of the polar nanoregions in terms of Pb-displacements only 
might be a too simple model. From single crystal 
data, we have only indirect evidence for displacements of cations at B and O sites.
These ions then probably play a less important role in the formation of diffuse scattering. 
 However, we note that qualitative understanding of x-ray 
diffuse scattering could be obtained by considering only Pb displacements. 
This might be due to the smaller x-ray cross-section of O compared with the metal cations. Neutron 
diffuse scattering analysis could then probably shed more light on O behavior 
if  the diffuse scattering was measured with a higher quality than what 
was possible up to now.
\item[iii)\quad ] The main structural feature causing the characteristic butterfly diffuse scattering in PMT 
is the formation of PNR where Pb atoms are displaced along one $<1,1,1>$ direction;
 PNR have large correlation lengths, except along the $<1,1,-2>$ direction orthogonal to the Pb displacement
direction, thereby confirming the 'pancake' model of ref.~\cite{xu2004}, but with a different geometry. 
\end{itemize}

In conclusion, we have analyzed the distribution of displacements in PMT by single crystal neutron 
diffraction and synchrotron radiation. The diffuse scattering and especially its structural origin have 
been understood qualitatively.
Another interesting topic is the relationship between coherent Pb $<111>$-displacements 
in PNRs and the known B-site partial ordering, with $(1/2,1/2,1/2)$ broad satellites (see Fig.~\ref{fig2}b) pointing to 
an ordering of $[111]$ B-site planes with alternating full-B'' (Ta) layers 
and 2/3-B',1/3-B'' layers. A comprehensive and detailed microstructural study 
of all systematic atomic displacements is necessary to understand this relationship and to 
bridge the gap between structure and macroscopic properties of relaxor ferroelectrics.
\section{Acknowledgment}
Work based on experiments performed at the Swiss-Norwegian Beamline SNBL and the Institute Laue-Langevin ILL of Grenoble, France. 
We thank D.~Chernyshov (SNBL) for his assistance with measurements of the diffuse scattering. 
RAC wishes to thank the Leverhulme trust for their financial support. This work was partially supported by 
the Swiss National Foundation (Project No. 20002-111545).
\newpage
\begin{table}
\caption{
Atomic parameters obtained for PMT at $T=20$~K. The atomic displacements are modeled with three split positions 
for the Pb ions (Pb$_x$: Pb displacements along $<1,0,0>$; Pb$_{xx}$ along $<1,1,0>$ and Pb$_{xxx}$ along $<1,1,1>$).
G is the number of ions in the unit cell; B$_{22}$=B$_{11}$, B$_{33}$ anisotropic and B$_{iso}$ 
isotropic temperature factors in (\AA$^2$).
\label{tab1}}
\begin{ruledtabular}
\begin{tabular}{lclclclclclclclc}
Atom&G&x&y&z&B$_{11}$&B$_{33}$&B$_{iso}$\\
\hline
20 K&&&&&&&\\
Pb$_{x}$&1/3&0.090(4)&0&0&-&-&0.88(7)\\
Pb$_{xx}$&1/3&0.056(4)&x&0&-&-&0.88(7)\\
Pb$_{xxx}$&1/3&0.045(3)&x&x&-&-&0.88(7)\\
O&3&1/2&1/2&0&1.86(4)&0.960(4)&\\
Ta/Mg&1&1/2&1/2&1/2&0.586(8)&B$_{11}$&\\
\hline
300 K&&&&&&&\\
Pb$_{x}$&1/3&0.076(9)&0&0&-&-&1.2(2)\\
Pb$_{xx}$&1/3&0.05(1)&x&0&-&-&1.2(2)\\
Pb$_{xxx}$&1/3&0.046(8)&x&x&-&-&1.2(2)\\
O&3&1/2&1/2&0&1.96(7)&0.85(7)&\\
Ta/Mg&1&1/2&1/2&1/2&0.62(2)&B$_{11}$&\\
\end{tabular}
\end{ruledtabular}
\end{table}

\begin{thebibliography}{99}
\bibitem{smolenskii59} G.A. Smolenskii and A.I. Agranovskaya, Soviet Physics Solid State {\bf 1}, 1429 (1959).
\bibitem{burns} G. Burns and B.A. Scott, Solid State Commun. {\bf 13}, 423 (1973) .
\bibitem{gvasaliya_pmn_prb} S.N. Gvasaliya, S.G. Lushnikov and B. Roessli, Phys. Rev. B {\bf 69}, 092105 (2004). 
\bibitem{hiraka} H. Hiraka, S.H. Lee, P.M. Gehring, G.Y. Xu and G. Shirane, Phys. Rev. B {\bf 70}, 184105 (2004).
\bibitem{gvasaliya_pmn_jpco} S.N. Gvasaliya, B. Roessli, R.A. Cowley, P. Huber and S.G. Lushnikov, J. Phys.: Condensed Matter {\bf 17}, 4343 (2005). 
\bibitem{cowley} R.A. Cowley, S.N. Gvasaliya and B. Roessli, Ferroelectrics {\bf 378}, 53 (2009).
\bibitem{vakhrushev95} S.B. Vakhrushev, A.A. Naberezhnov, N.M. Okuneva, and B.N. Savenko, 
Phys. Solid State {\bf 37}, 1993 (1995).
\bibitem{xu2004}G. Xu, Z. Zhong, H. Hiraka, and G. Shirane, Phys. Rev. B {\bf 70}, 174109 (2004).
\bibitem{welberry2006} T.R. Welberry, D.J. Goossens and M.J. Gutmann, Phys. Rev. B {\bf 74}, 224108 (2006).
\bibitem{bonneau} P. Bonneau, P. Garnier, G. Calvarin, E. Husson, J.R. Gavarri, A.W. Hewat and A. Morell, J. Solid State Chem. {\bf 91}, 350 (1991).
\bibitem{zhukov} V.V.~Chernyshev, S.G.~Zhukov, A.V.~Yatsenko, L.A.~Aslanov and H.~Schenk,
Acta Cryst. {\bf A50}, 601 (1994).     
\bibitem{vakhrushev2002} S.B. Vakhrushev and N.M. Okuneva, in Fundamental Physics of Ferroelectrics 2002, 
                         edited by Ronald E. Cohen, AIP Conf. Proc. {\bf 626} (AIP, Melville, NY, 2002), p. 117. 
\bibitem{markovin1992} O.Yu. Korshunov, P.A. Markovin, and R.V. Pisarev, Ferroelectrics {\bf 13}, 137 (1992 ).
\bibitem{gvasaliya2004} S.N. Gvasaliya, B. Roessli, D. Sheptyakov, S.G. Lushnikov and T.A. Shaplygina, 
                        Eur. Phys. J. B {\bf 40}, 235 (2004).
\bibitem{you} H. You and Q.M. Zhang, Phys. Rev. Lett. {\bf79}, 3950 (1997).
\bibitem{kirvoglaz} M.A. Krivoglaz, {\it Theory of X-ray and Thermal-Neutron Scattering by Real Crystals}, (Plenum, New-York, 1969).
\bibitem{gvasaliya2003} S.N. Gvasaliya, S.G. Lushnikov and B. Roessli, Europhys. Lett. {\bf 63}, 303 (2003).
\bibitem{gehring_2009} see Fig.2 in P. M. Gehring, H. Hiraka, C. Stock, 
         S.-H. Lee, W. Chen, Z.-G. Ye, S. B. Vakhrushev and Z. Chowdhuri, arXiv:0904.4234v1 [cond-matt.]
\bibitem{you2000} H. You, J. Phys. Chem. Solids {\bf 61}, 215 (2000).
\bibitem{antonio} If the autocorrelation function has cylindrical symmetry $G({\bf r})=\exp(-|\frac{{\bf w_1}\cdot{\bf r}}{L_1}|)\exp(-\frac{r_\perp}{L_r})$
with $L_1$ along the cylinder axis and $L_r$ perpendicular to it, the Fourier transforms is 
$\tilde{G}({\bf Q})=2\pi L^2_rL_1/{(1+(2\pi L_1{(\bf w_1}\cdot{\bf q})^2)}/{(1+(2\pi L_r q_\perp)^2)^{1.5}}$. 
\bibitem{pasciak2007} M. Pa{\'s}ciak, M. Wo{\l}cyrz and A. Pietraszko, Phys. Rev. B {\bf 76}, 014117 (2007).
\bibitem{jeong2005}I.-K. Jeong , T. W. Darling, J. K. Lee, Th. Proffen, R. H. Heffner, J. S. Park, K. S. Hong, W. Dmowski and T. Egami, Phys. Rev. Lett. {\bf 94}, 147602 (2005).
\bibitem{calvarin} G. Calvarin, E. Husson and Z.G. Ye, Ferroelectrics {\bf 165}, 349 (1995).
\bibitem{chetverikov} Y.O.~Chetverikov, A.A.~Naberezhnov, S.B.~Vakhrushev, B.~Dorner, and A.S.~Ivanov, 
Appl. Phys. A: Mater. Sci. Process. {\bf A74}, S989 (2002). 
\bibitem{li2008} Zhenrong Li, Zhuo Xu, Xi Yao and Z.-Y. Cheng, J. Appl. Phys. {\bf 104}, 024112 (2008).
\bibitem{giacovazzo} C. Giacovazzo, H.L. Monaco, D. Viterbo, F. Scordari, G. Gilli, G. Zanotti, and M. Catti,
"Fundamentals of Crystallography" (1992, Oxford: Oxford University Press).
\bibitem{jana} V. Petricek, M. Dusek, and L. Palatinus (2000). Jana2000. The crystallographic computing system. 
Institute of Physics, Praha, Czech Republic. 
\end{thebibliography}
\end{document}